\documentclass[prl,showpacs,twocolumn,superscriptaddress,floatfix,amsmath,raggedbottom,A4]{revtex4}
\usepackage[latin1]{inputenc}
\usepackage[english]{babel}
\usepackage{graphicx}
\usepackage{amssymb}
\usepackage{amsmath}
\usepackage{amscd}
\usepackage{eucal}
\usepackage{color}
\usepackage{bm}

\newcommand{\rmd}{{\rm d}}
\newcommand{\rmi}{{\rm i}}

\newcommand{\half}{{\textstyle{\frac{1}{2}}}}

\newcommand{\quarter}{\textstyle{\frac{1}{4}}}
\newcommand{\up}{\uparrow}
\newcommand{\dn}{\downarrow}

\newcommand{\al}{\alpha}

\newcommand{\eps}{\epsilon}

\newcommand{\Om}{\Omega}

\definecolor{DarkGreen}{rgb}{0,0.7,0}
\definecolor{Blue}{rgb}{0,0,0.8}
\definecolor{Red}{rgb}{1,0,0}

\begin{document}

\title{Temperature can enhance coherent oscillations at a Landau-Zener transition}
\author{Robert S.~Whitney}

\affiliation{
Laboratoire de Physique et Mod\'elisation des Milieux Condens\'es (UMR 5493), 
Universit\'e Joseph Fourier and CNRS, Maison des Magist\`eres, BP 166, 38042 Grenoble, France.}

\affiliation{Institut Laue-Langevin,
6 rue Jules Horowitz, BP 156, 38042 Grenoble, France.}

\author{Maxime Clusel}
\affiliation{Laboratoire Charles Coulomb, UMR 5221, CNRS and Universit\'e Montpellier 2, Montpellier, France. }

\affiliation{Institut Laue-Langevin,
6 rue Jules Horowitz, BP 156, 38042 Grenoble, France.}

\author{Timothy Ziman}

\affiliation{
Laboratoire de Physique et Mod\'elisation des Milieux Condens\'es (UMR 5493), 
Universit\'e Joseph Fourier and CNRS, Maison des Magist\`eres, BP 166, 38042 Grenoble, France.}

\affiliation{Institut Laue-Langevin,
6 rue Jules Horowitz, BP 156, 38042 Grenoble, France.}

\date{June 27, 2011}
\begin{abstract}
We consider sweeping a system through a Landau-Zener avoided-crossing, when that system is also coupled to an environment or noise. Unsurprisingly, we find that decoherence suppresses the coherent oscillations of quantum superpositions of system states, as superpositions decohere into mixed states. However, we also find an effect we call ``Lamb-assisted coherent oscillations'', in which a Lamb shift {\it exponentially enhances} the coherent oscillation amplitude. This dominates for high-frequency environments such as super-Ohmic environments, where the coherent oscillations can grow exponentially as either the environment coupling or temperature are {\it increased}. The  effect could be used as an experimental probe for high-frequency environments in such systems as molecular magnets, solid-state qubits, spin-polarized gases (neutrons or He3) or Bose-condensates.
\end{abstract}
\maketitle

{\bf Introduction.}
A central aspect of quantum systems is that they can be in a coherent superposition of different eigenstates, with observables then undergoing coherent oscillations.  One way to create such superpositions, is to take a system in its ground state and change its Hamiltonian too rapidly for the system to adiabatically follow the ground state. To distinguish between superpositions and incoherent mixtures, one needs to detect the relatively fast coherent oscillations.  These are now detectable in many systems, including superconducting qubits \cite{Rudner-et-al}, Bose-condensates \cite{Bose-condensates1,Bose-condensates2}, polarized He3 \cite{He3} or neutrons \cite{Rauchs-group},
and probably in molecular magnets \cite{Clusel-Ziman}. 

All quantum systems have some coupling to degrees of freedom in their environment. 
 Generally this coupling suppresses coherent oscillations as quantum superpositions decay into incoherent mixtures. The decay mechanism, called \textit{decoherence}, is usually stronger at higher environment temperature \cite{book:Breuer-Petruccione}.
In this letter we ask if this is always the case, by examining the archetypal example of the Landau-Zener transition, which generates a superposition of the ground and an excited state.  Various models of  the environment  will be considered: in some it behaves as a classical noise field, while in others its quantum nature  is taken into account.

We will arrive at the surprising conclusion that the environment can exponentially {\it enhance} the coherent oscillations generated at a Landau-Zener transition. This occurs because it modifies the coherent oscillations in two ways. The first is the standard decoherence mechanism, responsible for level-broadening,  
which suppresses the oscillations \cite{book:Breuer-Petruccione}. The second is a Lamb shift  of the levels   \cite{book:Cohen-Tannoudji} which can exponentially reduce or enhance the oscillations. To illustrate these effects,
we consider three types of environment: Markovian environments (which we will see exhibit decoherence only); High-frequency environments (which will exhibit Lamb shift only); Caldeira-Leggett sub- and super-Ohmic environments (exhibiting 
decoherence and Lamb shift). 
The last will show multiple regimes, due to competition between the two effects.

These noise-enhanced oscillations may remind one of quantum stochastic resonances (QSR)  
\cite{QSR-general,QSR-coloured}, 
however there are crucial differences. We have free coherent oscillations
at a frequency given by the level-splitting,  while QSR's driven oscillations are at the drive's frequency.  
QSR is typically an enhancement going like a power-law of dissipative rates,
with some non-exponential modification in those cases where the noise is coloured (and thus induces a Lamb-shift)
\cite{QSR-coloured}.
In contrast, the free oscillations we discuss are {\it exponentially} enhanced by the stochasticity 
due to an interplay of a Lamb shift and a Landau-Zener transition.

%
%
%
\begin{figure}
\includegraphics[width=8.5cm]{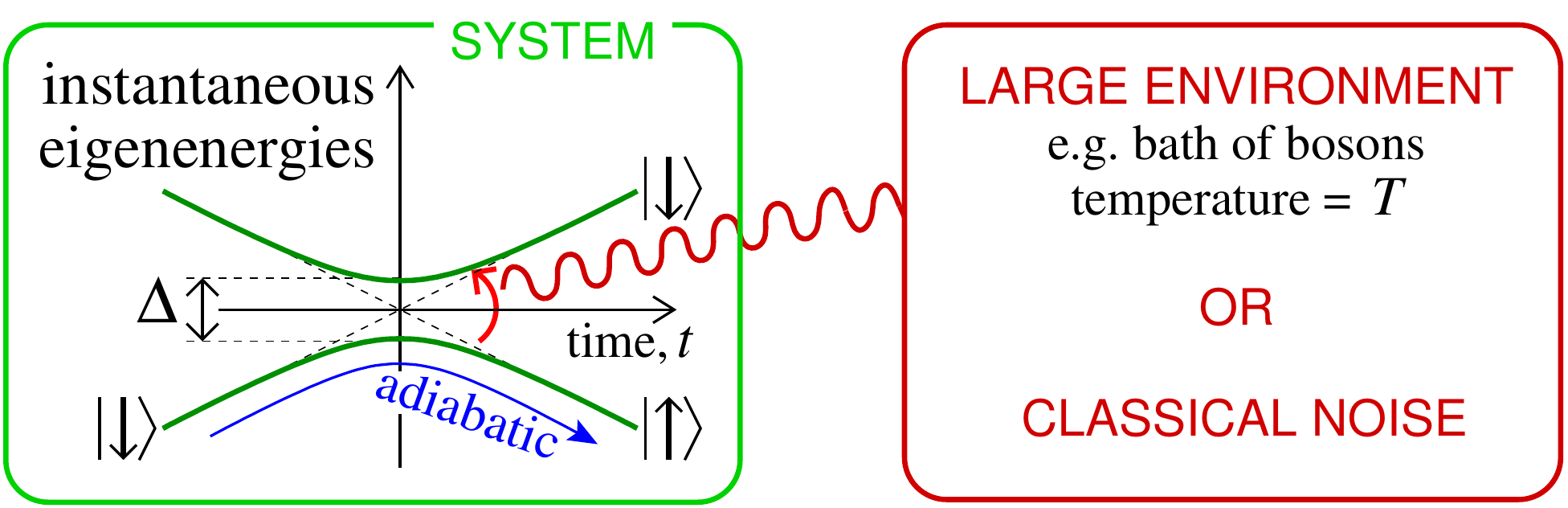}
\caption{\label{Fig:LZ} 
We ask how coupling to an environment (with a given spectrum) 
affects the coherent oscillations of a system swept through a Landau-Zener transition.
}
\end{figure}
%
%
%
\begin{figure*}
\includegraphics[width=\textwidth]{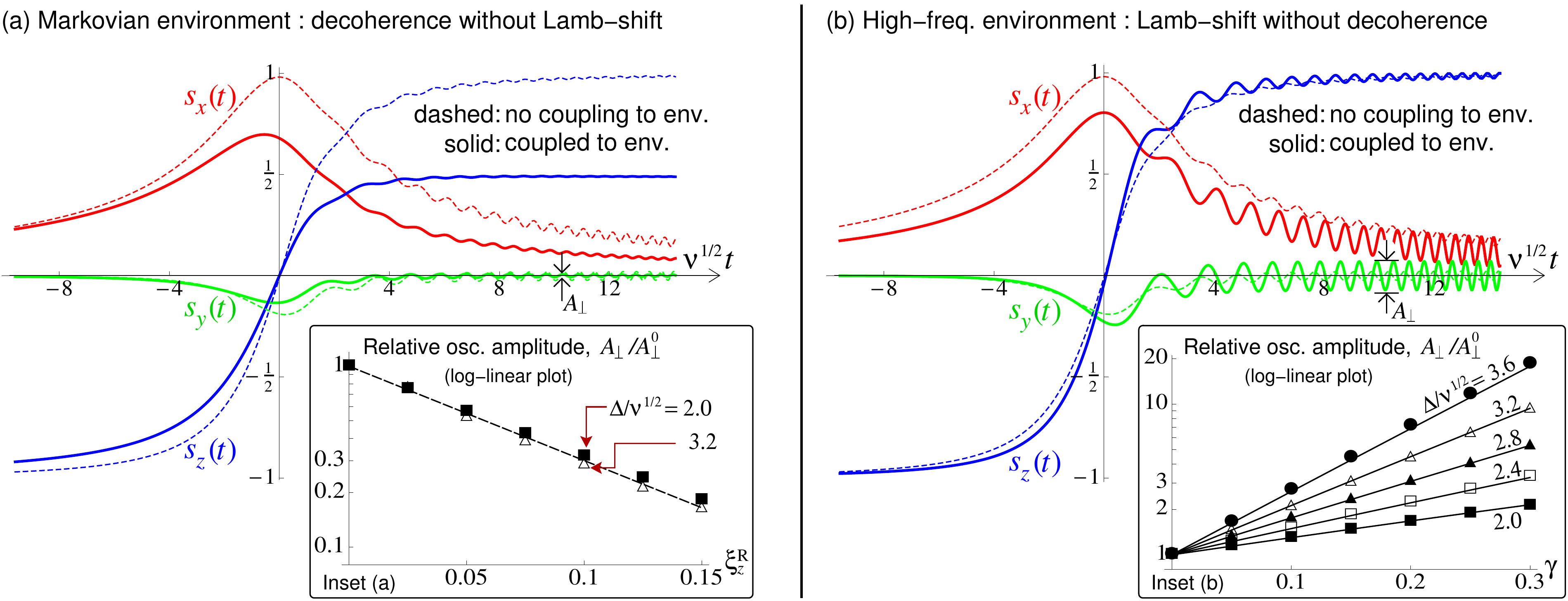}
\caption{\label{Fig:data} 
The three spin components of the system swept through a Landau-Zener transition,
given by numerical evolution of Eq.~(\ref{Eq:Bloch-Redfield-as-matrix}).
The main plots  in (a) and (b), have $\Delta/\nu^{1/2}=2.4$ in the presence (solid-curves) and absence (dashed-curves)
of environments.
In (a) the environment induces decoherence reducing the magnitude of the coherent oscillations, 
while in (b) the environment induces a Lamb shift  which \textit{enhances} such oscillations. 
 In (a) the only non-zero $\xi$ is $\xi^\mathrm{R}_z=0.1$. The oscillation magnitude $A_\perp$ is found by  
 fitting $s_y(t)$ for $\nu^{1/2}t$ from  9.5 to 10.5.  The inset shows its exponential decay with coupling-strength, 
 $\xi_z^\mathrm{R}$, at a rate which appears to be $\Delta$-independent.
In (b) the only non-zero $\xi$ is $\xi^\mathrm{R}_y=-\gamma\Delta$ with $\gamma=0.3$, for which Eq.~(\ref{Eq:Bloch-Redfield-as-matrix}) reduces to Eq.~(\ref{Eq:Bloch-Redfield-as-matrix-high-freq}). 
The inset shows $A_\perp$ growing exponentially with environment coupling, the solid-lines 
are the hypothesis that $A_\perp= (1-\gamma)^{1/2} A_\perp^0\big(\Delta (1-\gamma)^{1/2}\big)$ 
where $A_\perp^0(x)$ is given by Eq.~(\ref{Eq:A_perp-Zener}).
 Analysis of oscillations of $s_x(t)$ gives identical results.
}
\end{figure*}
%

{\bf Model.}
The Hamiltonian for the Landau-Zener transition of a two-level system is 
$\mathcal{H}^{\mathrm{LZ}}_t = -\half \big(\nu t \,{\bm \sigma}_z + \Delta \,{\bm \sigma}_x\big)$
when written in terms of Pauli matrices. The avoided-crossing occurs at time $t=0$, has width $\Delta$ and is swept through at rate $\nu$. We consider this system coupled to an environment (Fig.~\ref{Fig:LZ}), with the Hamiltonian
\begin{eqnarray}
\mathcal{H}^{\mathrm{sys\&env}}_t = \mathcal{ H}^{\mathrm{LZ}}_t -\half {\bm \sigma}_z {\bm X} + \mathcal{H}^{\mathrm{env}},
\label{Eq:H_sys+env}
\end{eqnarray}
where the environment operator ${\bm X}$ acts weakly on a huge number of environment modes. The environment Hamiltonian, $\mathcal{ H}^\mathrm{ env}$, is such that these modes have a broad, effectively continuous spread of frequencies.

Such models are well studied
for classical noise \cite{kayanuma1&2,Vestgarden-Galperin} and quantum environments (addressed using 
approximate \cite{Gefen-Benjacob-Caldeira1987,Shimshoni-Stern1993,Ao-Rammer,Pokrovsky-Sun}, 
exact \cite{Hanggi-Kayanuma},
or numerical methods \cite{Nalbach-Thorwart,LeHur}).
However they focus on the transition probability given by $\langle {\bm \sigma}_z(t)\rangle$. 
Now that one can measure $s_j(t) = \langle {\bm \sigma}_j(t)\rangle$ for $j=x,y,z$ in various systems 
\cite{Rudner-et-al,Bose-condensates1,He3,Rauchs-group,Clusel-Ziman}, we emphasis that they give us much more information than $s_z(t)$ alone. In this letter, we consider the magnitude of the coherent oscillations, $A_\perp$, defined by writing $s_y(t)=  \half A_\perp \sin \Phi_t$ for large times $t$.  Without an environment, Zener gives
\begin{eqnarray}
A_\perp^0(\Delta) &=&  4e^{-\pi \Delta^2/(4\nu)},  
\label{Eq:A_perp-Zener}
\end{eqnarray}
for a near adiabatic transition ($\Delta^2 \gg \nu$)\cite{Zener}. We will show how the environment could modify this relation.

{\bf Master equation.}
If the system-environment coupling in $\mathcal{H}^{\mathrm{sys\&env}}_t $
is weak enough to treat in a ``golden-rule'' manner, then
the spin's density matrix obeys the master equation
$\dot {\bm \rho}_t = -\rmi\left[\mathcal{ H}^\mathrm{ LZ}_t,{\bm \rho}_t\right]_-  - \quarter \big[{\bm \sigma}_z,(\Xi_t\,{\bm \rho}_t - {\bm \rho}_t\,\Xi^\dagger_t)\big]_-$, where the dot denotes a time-derivative \cite{master-eqns}.
The spin-operator $\Xi_{t}= \Xi^{\rm R}_t+ \rmi \Xi^{\rm I}_t$ is defined by
$\Xi_t ^{\rm R,I}= \int_0^\infty \rmd t_1 \al^{\rm R,I}(t_1) \, \mathcal{ U}_{t,t_1}\, {\bm \sigma}_z \,\mathcal{ U}^{-1}_{t,t_1}$, 
where  $\mathcal{ U}_{t,t_1}$ is the evolution under $\mathcal{ H}^\mathrm{ LZ}_\tau$ from time $(t-t_1)$ to time $t$.
We have split the environment correlation function
$\al(t) =\langle e^{\rmi \mathcal{ H}^\mathrm{ env}t} {\bm X}e^{-\rmi \mathcal{ H}^\mathrm{ env}t}{\bm X}\rangle$
into its real and imaginary parts, $\al_{\rm R}(t)$ and $\al_{\rm I}(t)$,
because they are the Fourier transforms of the environment's symmetric and antisymmetric spectral densities, 
$\mathcal{ S}(\Om)$ and $\mathcal{ A}(\Om)$. 
Any environment in equilibrium at temperature $T$ has $\mathcal{ A}(\Om)= \mathcal{ S}(\Om) \tanh[ \Om/(2k_\mathrm{ B}T)]$ \cite{whitney08}.
This master equation includes weak memory effects; it 
only reduces to Lindblad's markovian case if $\al(t)$ is a $\delta$-function \cite{whitney08}.

We parametrize
$
{\bm \rho}_t  = 
\half \left( 1+ s_x {\bm \sigma}_x +  s_y {\bm \sigma}_\mathrm{ y} + s_z {\bm \sigma}_\mathrm{ z}\right)$,
so that ${\bm s}=(s_x,s_y,s_z)$ is the spin-polarization vector, 
and  define ${\bm B}_t= (\Delta, 0,\nu t)$ and $ {\bm e}_z=(0,0,1)$, leading to 
\begin{eqnarray}
\dot {\bm s} = {\bm B}_t\times {\bm s} +  {\bm e}_z\times \left[{\bm \xi}^\mathrm{ R}(t)\times {\bm s}
\,+\,  {\bm \xi}^\mathrm{ I}(t) \right],
\label{Eq:Bloch-Redfield-as-matrix}
\end{eqnarray}
where 
$ \xi^\mathrm{ R,I}_j(t) $ are the ${\bm \sigma}_j$ components of   $\Xi^{\rm R,I}_t$.

Alternatively, Eq.~(\ref{Eq:Bloch-Redfield-as-matrix}) describes the  noise-averaged evolution under the Hamiltonian $\mathcal{ H}^{\rm LZ}_t - \half X_t \,{\bm \sigma}_z$,  with the noise-field $X_t$ treated using golden-rule \cite{Redfield}, with $\overline{X_t}=0$ and $\overline{X_{t}X_{t'}} = \al_{\rm R} (t-t')$. Then $\mathcal{ S}(\Om)$ is the noise-power at frequency $\Om$, while $\mathcal{ A}(\Om)=0$.

The correlation function $\al(t)$ typically decays on a timescale $\Om_\mathrm{ m}^{-1}$, where $\Om_\mathrm{ m}$ is the 
characteristic frequency of $\mathcal{ S}(\Om)$ and $\mathcal{ A}(\Om)$. We assume sufficiently fast decay ($\Om_\mathrm{ m}\gg\Delta,\nu^{1/2}$) that $\mathcal{ U}_{t,t_1} \simeq \exp[\rmi \mathcal{ H}_t  t_1]$ for all relevant $t_1$ in $\Xi_t$. 
Decoherence is due to 
$\xi^\mathrm{ R}_z = \big[\nu^2 t^2  \mathcal{ S}(0) \, +\, \Delta^2  \mathcal{ S}(B_t) \big]/(2B_t^2)$
and 
$\xi^\mathrm{ R}_x =\Delta \nu t \left[\mathcal{ S}(0) -\mathcal{ S}(B_t) \right]/(2B_t^2)$, for $B_t=|{\bm B}_t|$.
If these are much smaller than $B_t$, the decoherence rate 
$T_2^{-1}=\left[ (\Delta^2 + 2\nu^2t^2)\xi^\mathrm{ R}_z + \Delta \nu t \xi^\mathrm{ R}_x \right]/(2B_t^2)$
\cite{footnote:T_2}.
The Lamb shift is due to $\xi^\mathrm{ R}_y$. 
Defining $\gamma$ as the relative gap reduction due to this Lamb shift, 
we have
\begin{eqnarray}
\gamma \equiv -{\xi^\mathrm{ R}_y \over \Delta}  =  
\int_{-\infty}^\infty {\rmd \Om \over 2 \pi} \,{\mathcal{ S}(\Om) \over \Om^2 -B_t^2}.
\label{Eq:relative-shift}
\end{eqnarray}
Finally
$\xi^\mathrm{ I}_x = \Delta \nu t \int_{-\infty}^\infty (\rmd \Om/2 \pi) \,\mathcal{ A}(\Om)/(\Om^3 -B_t^2\Om)$
and
$\xi^\mathrm{ I}_y = \Delta\mathcal{ A}(B_t)/(2B_t)$.

\begin{figure}
\includegraphics[width=7.5cm]{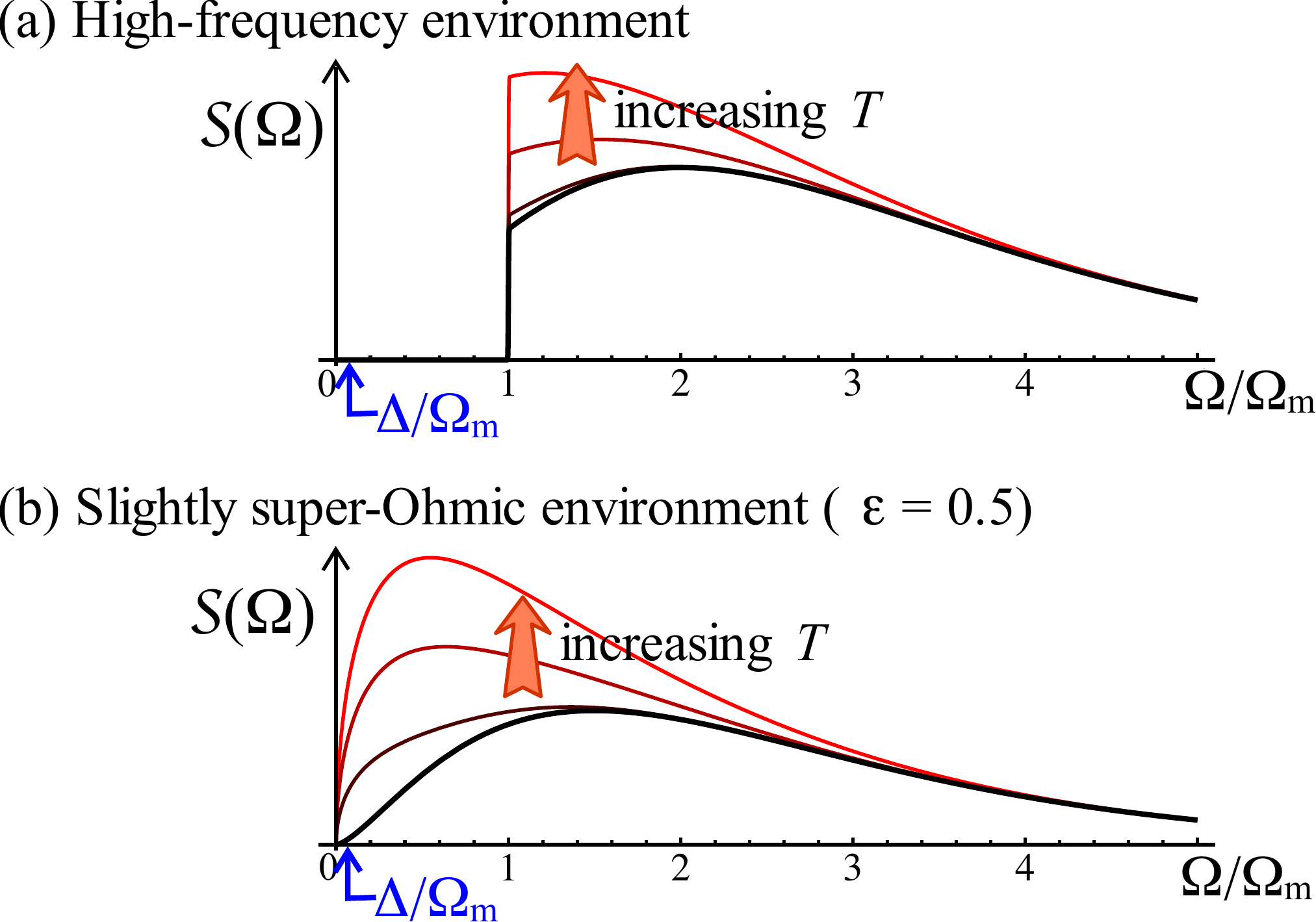}
\caption{ \label{Fig:S}
Plots of $\mathcal{S}(\Om)$ showing how weight below $k_\mathrm{B}T$ grows significantly as one increases $T$, while that above only grows slightly; the curves are for $k_\mathrm{B}T/\Om_\mathrm{m}=$ 0, 1/3, 2/3 and 1. In (a) there is no weight at low $\Om$, so increasing $T$ enhances $\gamma$ via the level-repulsion discussed in the text. For (b) there is enough weight at low $\Om$ that increasing $T$ reduces $\gamma$.}
\end{figure}

{\bf Markovian evolution.} We start with classical white-noise, with a noise-spectrum $\mathcal{ S}(\Om)$ that is 
$\Om-$independent (while  $\mathcal{ A}(\Om)=0$). The noise correlation function $\al(t) = a\delta(t)$, corresponds to a complete absence of memory. All $\xi$s in  Eq.~(\ref{Eq:Bloch-Redfield-as-matrix}) 
are zero except $\xi^\mathrm{ R}_z=\half {\cal S}(0)$,   so there is decoherence but no Lamb shift.
This gives the evolution in Fig.~\ref{Fig:data}a, with noise suppressing the oscillations.

{\bf High-frequency noise or environment.} Consider classical noise with $\mathcal{ S}(\Om)$
at much higher-frequencies than $B_t$. Only $\xi^\mathrm{ R}_y$ is non-zero, so there is a Lamb shift but no decoherence. Eq.~(\ref{Eq:Bloch-Redfield-as-matrix}) reduces to
\begin{eqnarray}
\left(\!\begin{array}{c}  \dot s_x \\  \dot s_y  \\  \dot s_z \end{array}\!\right) 
= \left(\!\begin{array}{ccc} 0 & \nu t & 0  \\  
-\nu t& 0 & \Delta (1-\gamma)  \\  0 & -\Delta & 0\end{array}\!\right)
\left(\!\begin{array}{c}  s_x \\  s_y  \\  s_z \end{array}\!\right). 
\label{Eq:Bloch-Redfield-as-matrix-high-freq}
\end{eqnarray}
whose evolution is shown in Fig.~\ref{Fig:data}b.
The relative gap reduction  
$\gamma =c\, \mathcal{S}(\Om_\mathrm{m})/ \Om_\mathrm{ m}$,
with constant $c\sim {\cal O}[1]$.

For a quantum environment with $\mathcal{ S}(\Om)$ at much higher-frequencies than $B_t$ (as in Fig.~\ref{Fig:S}a),
only $\xi^\mathrm{ R}_y$ and $\xi^\mathrm{ I}_x$ are non-zero. For large $\Om_\mathrm{ m}$, we also have $\xi^\mathrm{ I}_x \ll  \xi^\mathrm{ R}_y$. Ignoring $\xi^\mathrm{ I}_x$, we recover  Eq.~(\ref{Eq:Bloch-Redfield-as-matrix-high-freq}), with $\gamma$ now depending on the environment temperature, $T$. For a spin-boson model ---
with ${\bm X}= \sum_n C({\bm a}^\dagger_n+{\bm a}_n)$ where ${\bm a}^\dagger_n$ (${\bm a}_n$) is the $n$th oscillator's creation (annihilation) operator --- we have $\mathcal{ S}(\Om)=2\pi C^2 d\big( |\Om| \big) \coth[|\Om|/(2k_\mathrm{ B}T)]$, where $d(\Om)$ is the oscillator density at $\Om$. So $\gamma$ is an {\it increasing} function of $T$, 
thus the coherent-oscillation magnitude, $A_\perp$, grows  {\it exponentially} 
with the environment coupling and strongly with its temperature. 
\textit{Exponential} growth with $T$ occurs for $\gamma\propto T$ when $k_{\rm B}T$ is larger than the typical $\Om$.

Fig.~\ref{Fig:S}a shows $d(\Om) = 
Ne^{-(\Omega/\Omega_{\rm m}-1)}/ (5\Om_\mathrm{ m}^3)$  
 for $\Om \geq \Om_\mathrm{ m}$ and zero elsewhere, with integrated density $N$.  
 Then  $\gamma$ is given by 
$\gamma 
= (K/ \Om_\mathrm{ m}) g\big({2k_\mathrm{ B}T\over \Om_\mathrm{ m}}\big)$, with $K=NC^2/\Om_\mathrm{ m}$ and 
$g(x)=\int_1^\infty {\rm d} \mu$ $\times e^{-\mu} {\rm coth}[\mu/x]$; so $g(x\ll 1)=1$ and $g(x \gg 1) \propto x$.

\begin{figure}
\includegraphics[width=8.0cm]{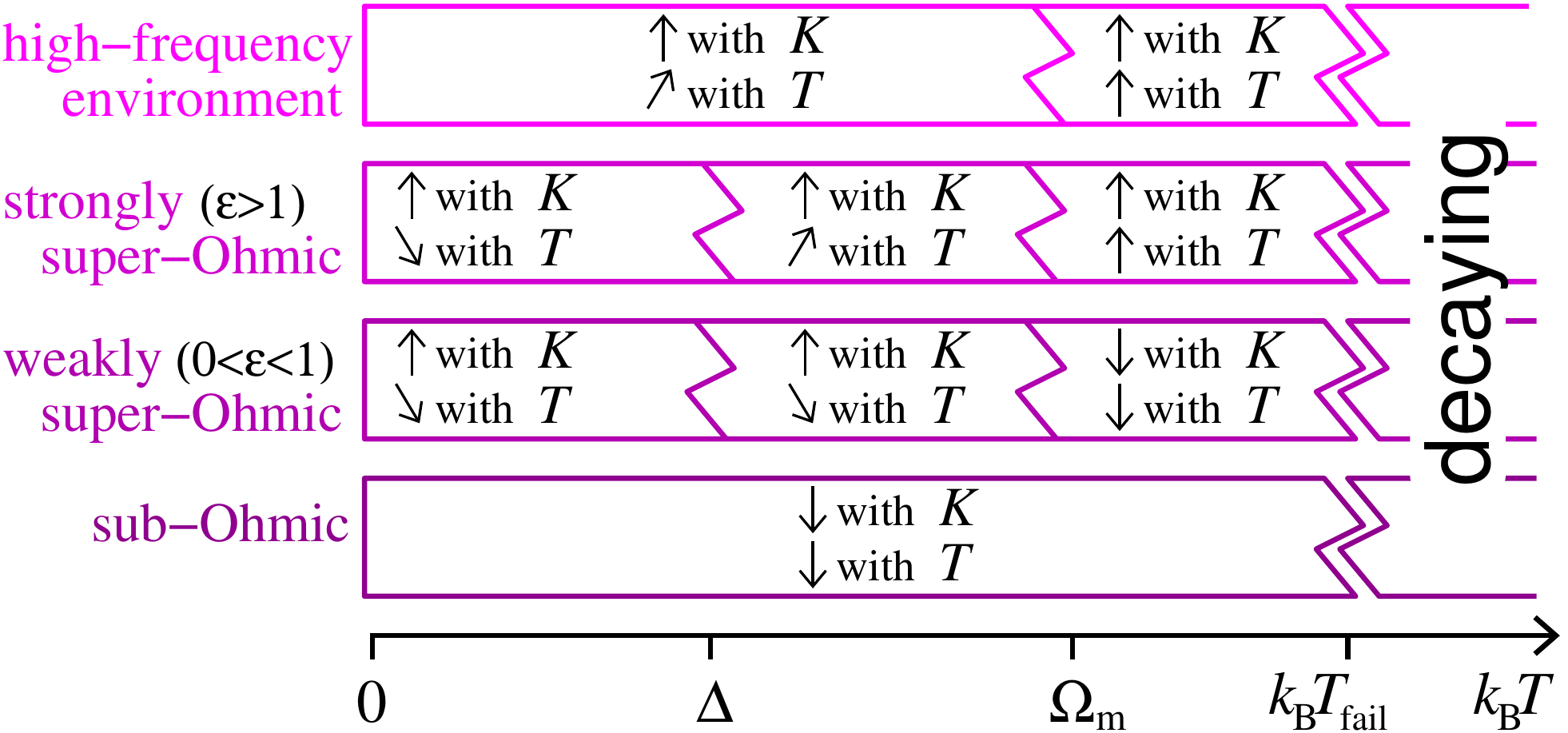}
\caption{ \label{Fig:regimes}
Regimes of behaviour of the coherent-oscillation magnitude, $A_\perp$.
Exponentially strong enhancement with increasing coupling is indicated by \hbox{``$\big\up \!$ with $\! K$''}
(and exponential reduction by \hbox{``$\big\dn\!$ with $\! K$''}). 
Exponentially strong increases and decreases  with increasing temperature are indicated
by  \hbox{``$\big\up \!$ with $\!  T$''} and  \hbox{``$\big\dn \!$ with $\!  T$''},
while weaker temperature dependences are indicated by  
\hbox{``$\nearrow \!$ with $\! T$''} and  \hbox{``$\searrow \!$ with $\! T$''}. }
\end{figure}

{\bf Sub- and super-ohmic environments.}
In this case all $\xi$s are finite, so both 
the Lamb shift  and decoherence are present. 
We take
$\mathcal{ S}(\Om) = K \big(|\Om|/\Om_\mathrm{ m}\big)^{1+\eps} 
e^{-|\Om|/\Om_\mathrm{ m}}\coth \left[{|\Om| /(2k_\mathrm{ B}T)}\right]$,
while $\mathcal{ A}(\Om)= \mathcal{ S}(\Om) \tanh\left[{\Om /(2k_\mathrm{ B}T)}\right]$. For $\eps <0$ this is \textit{sub-Ohmic}, while for $\eps >0$ it is \textit{super-Ohmic} \cite{Caldeira-Leggett-etc}; we assume that $\Om_\mathrm{ m}$ is very large. For a spin-boson model, this requires an oscillator density of $d(\Om)=N \big(\Om_\mathrm{ m}\Gamma(2+\eps)\big)^{-1}(\Om/\Om_\mathrm{ m})^{1+\eps} e^{-\Om/\Om_\mathrm{ m}}$, with $K=NC^2/(\Om_\mathrm{ m}\Gamma(2+\eps))$.
For sub-Ohmic cases ($\eps <0$), coherent oscillations are suppressed by both decoherence and a Lamb shift which increases the gap. 
For super-Ohmic cases ($\eps>0$) as in Fig.~\ref{Fig:S}b, 
the relative gap reduction due to the Lamb shift, $\gamma$,  has three regimes of behaviour: 
\begin{eqnarray}
\gamma 
= { K \over \Om_\mathrm{ m}}  \! \times \! \left\{ \!
\begin{array}{ccc}
\mu_0 - \mu^\mathrm{ low}_1 {\displaystyle {(k_\mathrm{ B}T)^{2+\eps} \over  \Delta^2 \Om_\mathrm{ m}^\eps}},   
& & \hskip -4mm  k_\mathrm{ B}T\ll \Delta \ll \Om_\mathrm{ m},
\\
\mu_0 \!+ \! \mu^\mathrm{ med}_1\, (k_\mathrm{ B}T/\Om_\mathrm{ m})^\eps,   
& 
{\phantom{\Bigg[  }}
& \hskip -4mm \Delta \ll k_\mathrm{ B}T\ll \Om_\mathrm{ m},
\\
\mu^\mathrm{ high}_1\, k_\mathrm{ B}T/\Om_\mathrm{ m},
& &\hskip -4mm \Delta \ll \Om_\mathrm{ m} \ll  k_\mathrm{ B}T.
\end{array}\right.
\label{Eq:all-regimes}
\end{eqnarray}
In each regime, we have split the contribution into a dominant $T$-independent part, $\mu_0$, and a $T$-dependent part (with prefactor $\mu_1$), subdominant except for high $T$.  
All $\mu$s are  $\mathcal{ O}[1]$; 
$\mu_0,\mu_1^\mathrm{ low}$ are negative for all $\eps>0$, 
while $\mu_1^\mathrm{ med},\mu_1^\mathrm{ high}$ are negative for $\eps<1$ and positive for $\eps >1$.
So if $k_{\rm B}T \gg \Delta$, $\gamma$ decays with $T$ for $\eps<1$ and grows for $\eps>1$.
At very long times, decoherence suppresses the oscillations. However for finite times, as in  Fig.~\ref{Fig:data}, comparing the  Lamb and decoherence effects   gives  Fig.~\ref{Fig:regimes}.

{\bf Beyond golden-rule.}
Using real-time Dyson equations to estimate corrections to the golden-rule \cite{whitney08}
one finds that Eq.~(\ref{Eq:Bloch-Redfield-as-matrix}) is valid for
$A \tau_{\rm corr}\ll1$, where $\tau_{\rm corr}$  is the time for $\alpha(t)$ to decay
and  $A=\int_0^\infty {\rm d}t\, \alpha(t)$.
Thus the golden-rule fails when $\gamma$ is not small; our estimates
indicate that oscillations then decay  (``decay'' in Fig.~\ref{Fig:regimes}).
If $K \ll \Om_\mathrm{ m}$, this occurs in the high-$T$ regime at 
$k_\mathrm{ B}T_\mathrm{fail}\sim\Om_\mathrm{ m}^2/K$.
If $K \gtrsim \Om_\mathrm{ m}$, there is no golden-rule regime
and oscillations are always suppressed.

{\bf Physical interpretation.}
The Lamb shift is due to level-repulsion between the spin and the environment modes: high-frequency modes reduce the gap in the spin-Hamiltonian, while low-frequency modes enhance it, see Eq.~(\ref{Eq:relative-shift}). The Landau-Zener transition is exponentially sensitive to this gap, thus a tiny reduction of it makes the transition much less adiabatic, so coherent oscillations are much larger. In contrast, decoherence comes from environment modes at zero-frequency or frequencies in resonance with the spin's level-spacing (see $\xi^\mathrm{R}_{x}$ and $\xi^\mathrm{R}_{z}$). As temperature grows, low-frequency modes are more enhanced than high-frequency ones (Fig.~\ref{Fig:S}); their competition causes the $T$-dependences in  Fig.~\ref{Fig:regimes}.

This picture neglects that the  Lamb shift occurs for only one $\Delta$-term in Eq.~(\ref{Eq:Bloch-Redfield-as-matrix-high-freq}), thereby modifying the nature of the dynamics (not just the gap). However Fig.~\ref{Fig:data}b confirms that the picture is qualitatively correct.  Thus we also expect that interference patterns due to multiple passages though an avoided-crossing
(Landau-Zener-St\"uckelberg interference) 
\cite{Landau-Zener-Stuckelberg-review,Landau-Zener-Stuckelberg1,Landau-Zener-Stuckelberg2,Nanomechanical} 
will grow exponentially with 
increasing $T$ whenever the environment is dominated by  high-frequencies.

{\bf Experimental applications.}
These ``Lamb-assisted coherent oscillations'' could be 
used to 
probe whether a system 
has a high-frequency environment (just as spin-echo is a probe for low-frequency environments \cite{qubit-spin-echo-expt}).
If the system Hamiltonian is static, 
then the Lamb shift only gives a weak $T$-dependence to the Larmor precession rate. 
However if one sweeps it through a Landau-Zener transition, 
the coherent oscillation magnitude becomes exponentially sensitive to
this $T$-dependent shift.
A potential application of this probe would be molecular magnets, 
where there are believed to be 
two potential sources of relaxation:
the primarily high-frequency bath of phonons \cite{footnote:phonon-bath},  and the low-frequency bath of nuclear spins.
Looking for ``Lamb-assisted coherent oscillations'' could clarify which dominates.


One could equally investigate whether high-frequency environments are important sources of dissipation in 
quantum systems as varied as superconducting qubits \cite{Rudner-et-al},
Bose-condensates \cite{Bose-condensates1,Bose-condensates2}, 
nano-mechanical resonators \cite{Nanomechanical},
or spin-polarized gases of He3 \cite{He3} or neutrons \cite{Rauchs-group}.




\begin{thebibliography}{1}
\bibitem{master-eqns}
This is the Bloch-Redfield equation \cite{Makhlin-review03,whitney08},
or the weak-coupling limit of the Nakajima-Zwanzig equation \cite{book:Breuer-Petruccione}.
\bibitem{footnote:phonon-bath}
Spins in each molecular-magnet most likely couple strongly to
optical phonons, they have energies of tens of meV, 
much higher than the typical $\Delta \sim 0.01-0.1$meV.
\end{thebibliography}
\end{document}